\documentclass[12pt]{article} \textwidth=161mm \textheight=223mm
\usepackage{graphicx}
\usepackage{amsbsy}
\usepackage{amsmath}
\usepackage{amsfonts}

\numberwithin{equation}{section}

\newtheorem{theorem}{Theorem}[section]
\newtheorem{corollary}{Corollary}[section]

\title{Computation of multiple eigenvalues and
generalized eigenvectors for matrices dependent on parameters}

\author{Alexei A. Mailybaev\footnote{Institute of Mechanics, Lomonosov Moscow State
University, Michurinsky pr. 1, 119192 Moscow, Russia. E-mail:
mailybaev@imec.msu.ru, Tel.: (7095) 939 2039, Fax: (7095) 939
0165.}}

\date{}

\begin{document}

\maketitle

\begin{abstract}
The paper develops Newton's method of finding multiple eigenvalues
with one Jordan block and corresponding generalized eigenvectors
for matrices dependent on parameters. It computes the nearest
value of a parameter vector with a matrix having a multiple
eigenvalue of given multiplicity. The method also works in the
whole matrix space (in the absence of parameters). The approach is
based on the versal deformation theory for matrices. Numerical
examples are given.
\end{abstract}

\noindent\textbf{Keywords:} multiparameter matrix family, multiple
eigenvalue, generalized eigenvector, Jordan block, versal
deformation, Schur decomposition

\section{Introduction}

Transformation of a square nonsymmetric (non-Hermitian) matrix
$\mathbf{A}$ to the Jordan canonical form is the classical subject
that finds various applications in pure and applied mathematics
and natural sciences. It is well known that a generic matrix has
only simple eigenvalues and its Jordan canonical form is a
diagonal matrix. Nevertheless, multiple eigenvalues typically
appear in matrix families, and one Jordan block is the most
typical Jordan structure of a multiple eigenvalue~\cite{Ar1, Ar2}.
Many interesting and important phenomena associated with
qualitative changes in the dynamics of mechanical
systems~\cite{Kirillov2005, MaSe1, MaSe2, SePe}, stability
optimization~\cite{BuLeOv, KiSe, Ov}, and bifurcations of
eigenvalues under matrix perturbations~\cite{MoBuOv,
SeyranianMailybaev2003, SeyranianEtAl2005, ViLy} are related to
multiple eigenvalues. Recently, multiple eigenvalues with one
Jordan block became of great interest in physics, including
quantum mechanics and nuclear physics~\cite{AntoniouEtAl2001,
Heiss2004b, LatinneEtAl1995}, optics~\cite{BerryDennis2003}, and
electrical engineering~\cite{Dobson2001}. In most applications,
multiple eigenvalues appear through the introduction of
parameters.

In the presence of multiple eigenvalues, the numerical problem of
computation of the Jordan canonical form is unstable, since the
degenerate structure can be destroyed by arbitrarily small
perturbations (caused, for example, by round-off errors). Hence,
instead of analyzing a single matrix, we should consider this
problem in some neighborhood in matrix or parameter space. Such
formulation leads to the important problem left open by
Wilkinson~\cite{Wi2, Wi3}: to find the distance of a given matrix
to the nearest degenerate matrix.

We study the problem of finding multiple eigenvalues for matrices
dependent on several parameters. This implies that matrix
perturbations are restricted to a specific submanifold in matrix
space. Such restriction is the main difficulty and difference of
this problem from the classical analysis in matrix spaces.
Existing approaches for finding matrices with multiple
eigenvalues~\cite{De, EdEl1, ElJoKa1, ElJoKa2, GoWi, KaRu, KaRu2,
Ku, LiEd, Ru, Wi2, Wi3} assume arbitrary perturbations of a matrix
and, hence, they do not work for multiparameter matrix families.
We also mention the topological method for the localization of
double eigenvalues in two-parameter matrix
families~\cite{KorschMossmann2003}.

In this paper, we develop Newton's method for finding multiple
eigenvalues with one Jordan block and corresponding generalized
eigenvectors in multiparameter matrix families. The presented
method solves numerically the Wilkinson problem of finding the
nearest matrix with a multiple eigenvalue (both in multiparameter
and matrix space formulations). The implementation of the method
in MATLAB code is available, see~\cite{MATLABcode}. The method is
based on the versal deformation theory for matrices. In spirit,
our approach is similar to~\cite{Fa}, where matrices with multiple
eigenvalues where found by path-following in matrix space
(multiparameter case was not considered).

The paper is organized as follows. In Section 2, we introduce
concepts of singularity theory and describe a general idea of the
paper. Section 3 provides expressions for values and derivatives
of versal deformation functions, which are used in Newton's method
in Section 4. Section 5 contains examples. In Section 6 we discuss
convergence and accuracy of the method. Section 7 analyzes the
relation of multiple eigenvalues with sensitivities of simple
eigenvalues of perturbed matrices. In Conclusion, we summarize
this contribution and discuss possible extensions of the method.
Proofs are collected in the Appendix.

\section{Multiple eigenvalues with one Jordan block
in multiparameter matrix families}

Let us consider an $m \times m$ complex non-Hermitian matrix
$\mathbf{A}$, which is an analytical function of a vector of
complex parameters $\mathbf{p} = (p_1,\ldots,p_n)$. Similarly, one
can consider real or complex matrices smoothly dependent on real
parameters, and we will comment the difference among these cases
where appropriate. Our goal is to find the values of parameter
vector $\mathbf{p}$ at which the matrix $\mathbf{A}(\mathbf{p})$
has an eigenvalue $\lambda$ of algebraic multiplicity $d$ with one
$d \times d$ Jordan block (geometric multiplicity $1$). Such and
eigenvalue $\lambda$ is called nonderogatory. There is a Jordan
chain of generalized vectors $\mathbf{u}_1,\ldots,\mathbf{u}_d$
(the eigenvector and associated vectors) corresponding to
$\lambda$ and determined by the equations
    \begin{equation}
    \label{3.1}
    \begin{array}{rcl}
        \mathbf{A}\mathbf{u}_1&=&\lambda\mathbf{u}_1,\\[3pt]
        \mathbf{A}\mathbf{u}_2&=&\lambda\mathbf{u}_2
        +\mathbf{u}_1,\\[3pt]
        &\vdots& \\[3pt]
        \mathbf{A}\mathbf{u}_d&=&\lambda\mathbf{u}_d+\mathbf{u}_{d-1}.
    \end{array}
    \end{equation}
These vectors form an $m \times d$ matrix $\mathbf{U} =
[\mathbf{u}_1,\ldots,\mathbf{u}_d]$ satisfying the equation
  \begin{equation}
  \label{3.2}
    \mathbf{A}\mathbf{U}=\mathbf{U}\mathbf{J}_\lambda,\qquad
    \mathbf{J}_\lambda=\left(\begin{array}{cccc}
      \lambda&1&&\\ &\lambda&\ddots&\\ &&\ddots&1\\ &&&\lambda
    \end{array}\right),
  \end{equation}
where $\mathbf{J}_\lambda$ is the Jordan block of size $d$. Recall
that the Jordan chains taken for all the eigenvalues and Jordan
blocks determine the transformation of the matrix $\mathbf{A}$ to
the Jordan canonical form~\cite{Gantmacher1998}.

\begin{figure}
\begin{center}
\includegraphics[angle=0, width=0.6\textwidth]{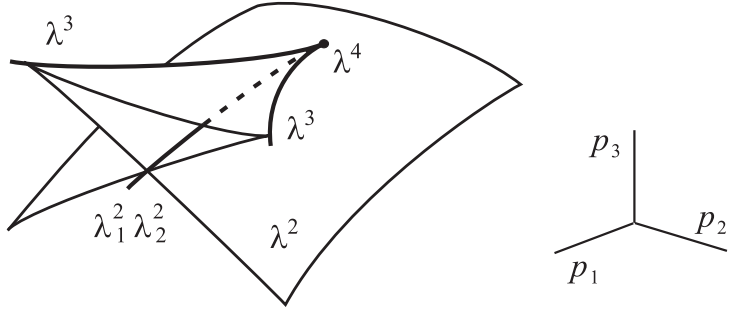}
\end{center}
\caption{Geometry of the bifurcation diagram.} \label{fig1}
\end{figure}

In singularity theory~\cite{Ar2}, parameter space is divided into
a set of strata (smooth submanifolds of different dimensions),
which correspond to different Jordan structures of the matrix
$\mathbf{A}$. Consider, for example the matrix family
\begin{equation}
  \mathbf{A}(\mathbf{p})=\left(\begin{array}{cccc}
    0 & 1 & 0 & 0 \\
    p_1 & 0 & 1 & 0 \\
    p_2 & 0 & 0 & 1 \\
    p_3 & 0 & 0 & 0
  \end{array}\right),\quad
  \mathbf{p}=(p_1,p_2,p_3).
\label{2.2}
\end{equation}
The bifurcation diagram in parameter space is shown in
Figure~\ref{fig1} (for simplicity, we consider only real values of
parameters). There are four degenerate strata: $\lambda^2$
(surfaces), $\lambda^3$ and $\lambda_1^2\lambda_2^2$ (curves), and
$\lambda^4$ (a point). The surface $\lambda^2$, curve $\lambda^3$,
and point $\lambda^4$ correspond, respectively, to the matrices
with double, triple, and quadruple eigenvalues with one Jordan
block. The curve $\lambda_1^2 \lambda_2^2$ is the transversal
self-intersection of the stratum $\lambda^2$ corresponding to the
matrices having two different double eigenvalues. This bifurcation
diagram represents the well-known ``swallow tail''
singularity~\cite{Ar2}.

We study the set of parameter vectors, denoted by $\lambda^d$,
corresponding to matrices having multiple eigenvalues with one
Jordan block of size $d$. The set $\lambda^d$ is a smooth surface
in parameter space having codimension $d-1$~\cite{Ar1, Ar2}. Thus,
the problem of finding multiple eigenvalues in a matrix family is
equivalent to finding the surface $\lambda^d$ or its particular
point. Since the surface $\lambda^d$ is smooth, we can find it
numerically by using Newton's method. This requires describing the
surface $\lambda^d$ as a solution of $d-1$ equations
    \begin{equation}
    q_i(\mathbf{p}) = 0, \qquad i = 2,\ldots,d,
    \label{2.3}
    \end{equation}
for independent smooth functions $q_i(\mathbf{p})$. (In these
notations, we keep the first function for the multiple eigenvalue
$\lambda = q_1(\mathbf{p})$.) Finding the functions
$q_i(\mathbf{p})$ and their first derivatives is the clue to the
problem solution.

In this paper, we define the functions $q_i(\mathbf{p})$ in the
following way. According to versal deformation theory~\cite{Ar1,
Ar2}, in the neighborhood of $\lambda^d$, the matrix
$\mathbf{A}(\mathbf{p})$ satisfies the relation
    \begin{equation}
    \mathbf{A}(\mathbf{p})\mathbf{U}(\mathbf{p})
    = \mathbf{U}(\mathbf{p})\mathbf{B}(\mathbf{p}),\quad
    \mathbf{B}(\mathbf{p})
    =\left(\begin{array}{cccc}
      q_1(\mathbf{p}) & 1 & & \\
      q_2(\mathbf{p})& q_1(\mathbf{p}) & \ddots & \\
      \vdots & & \ddots &1 \\
      q_d(\mathbf{p})& & & q_1(\mathbf{p})
    \end{array}\right),
    \label{2.4}
    \end{equation}
where $\mathbf{U}(\mathbf{p})$ is an $m \times d$ analytic matrix
family, and $q_1(\mathbf{p}),\ldots,q_d(\mathbf{p})$ are analytic
functions (blank places in the matrix are zeros). The functions
$q_1(\mathbf{p}),\ldots,q_d(\mathbf{p})$ are uniquely determined
by the matrix family $\mathbf{A}(\mathbf{p})$.

By using (\ref{2.4}), it is straightforward to see that the
surface $\lambda^d$ is defined by equations (\ref{2.3}). If
(\ref{2.3}) are satisfied, the matrix $\mathbf{B}(\mathbf{p})$ is
the $d \times d$ Jordan block. Hence, at $\mathbf{p} \in
\lambda^d$, the multiple eigenvalue is $\lambda = q_1(\mathbf{p})$
and the columns of $\mathbf{U}(\mathbf{p})$ are the generalized
eigenvectors satisfying equations (\ref{3.1}). The method of
finding the functions $q_i(\mathbf{p})$ and
$\mathbf{U}(\mathbf{p})$ and their derivatives at the point
$\mathbf{p} \in \lambda^d$ has been developed in \cite{Ma, Ma2}.
In Newton's method for solving (\ref{2.3}), we need the values and
derivatives of the functions $q_i(\mathbf{p})$ at an arbitrary
point $\mathbf{p} \notin \lambda^d$.

\section{Linearization of versal deformation functions}

Let $\mathbf{p}_0$ be a given parameter vector determining a
matrix $\mathbf{A}_0=\mathbf{A}(\mathbf{p}_0)$. Since multiple
eigenvalues are nongeneric, we typically deal with a
diagonalizable matrix $\mathbf{A}_0$. Let
$\lambda_1,\ldots,\lambda_m$ be eigenvalues of the matrix
$\mathbf{A}_0$. We sort these eigenvalues so that the first $d$ of
them, $\lambda_1,\ldots,\lambda_d$, coalesce as the parameter
vector is transferred continuously to the surface $\lambda^d$. The
eigenvalues that form a multiple eigenvalue are usually known from
the context of a particular problem. Otherwise, one can test
different sets of $d$ eigenvalues.

Let us choose $m\times d$ matrices $\mathbf{X}$ and $\mathbf{Y}$
such that
\begin{equation}
  \label{3.5a}
    \mathbf{A}_0\mathbf{X} = \mathbf{X}\mathbf{S}, \quad
     \mathbf{Y}^*\mathbf{A}_0 = \mathbf{S}\mathbf{Y}^*,
     \quad \mathbf{Y}^*\mathbf{X} = \mathbf{I},
  \end{equation}
where $\mathbf{S}$ is the $d \times d$ matrix whose eigenvalues
are $\lambda_1,\ldots,\lambda_d$; the star denotes the complex
conjugate transpose. The first two equalities in (\ref{3.5a})
imply that the columns of the matrix $\mathbf{X}$ span the right
invariant subspace of $\mathbf{A}_0$ corresponding to
$\lambda_1,\ldots,\lambda_d$, and the columns of $\mathbf{Y}$ span
the left invariant subspace. The third equality is the
normalization condition. The matrix $\mathbf{S}$ can be expressed
as
  \begin{equation}
  \label{3.5}
    \mathbf{S} = \mathbf{Y}^*\mathbf{A}_0\mathbf{X},
  \end{equation}
which means that $\mathbf{S}$ is the restriction of the matrix
operator $\mathbf{A}_0$ to the invariant subspace given by the
columns of $\mathbf{X}$. The constructive way of choosing the
matrices $\mathbf{S}$, $\mathbf{X}$, and $\mathbf{Y}$ will be
described in the next section.

The following theorem provides the values and derivatives of the
functions $q_1(\mathbf{p}),\ldots,q_d(\mathbf{p})$ in the versal
deformation (\ref{2.4}) at the point $\mathbf{p}_0$.

\begin{theorem}
\label{t1} Let $\mathbf{S}$, $\mathbf{Y}$, and $\mathbf{X}$ be the
matrices satisfying equations (\ref{3.5a}). Then
    \begin{equation}
    \label{3.6}
    q_1(\mathbf{p}_0) = \mathrm{trace}\,\mathbf{S}/d,
    \end{equation}
and the values of $q_2(\mathbf{p}_0),\ldots,q_d(\mathbf{p}_0)$ are
found as the characteristic polynomial coefficients of the
traceless matrix $\mathbf{S}-q_1(\mathbf{p}_0)\mathbf{I}$:
  \begin{equation}
  \label{3.8}
    z^d-q_2(\mathbf{p}_0)z^{d-2}-\cdots
    -q_{d-1}(\mathbf{p}_0)z-q_d(\mathbf{p}_0)
    =\det\big((z+q_1(\mathbf{p}_0))\mathbf{I}-\mathbf{S}\big),
  \end{equation}
where $\mathbf{I}$ is the $d \times d$ identity matrix. The first
derivatives of the functions $q_i(\mathbf{p})$ at $\mathbf{p}_0$
are determined by the recurrent formulae
  \begin{equation}
  \label{3.10}
    \begin{array}{l}
      \displaystyle
      \frac{\partial q_1}{\partial p_j}
      = \mathrm{trace}\left(\mathbf{Y}^
      *\frac{\partial\mathbf{A}}{\partial
      p_j}\mathbf{X}\right)/d, \\[10pt]
      \displaystyle
      \frac{\partial q_i}{\partial p_j}
      = \mathrm{trace}\left((\mathbf{S}-q_1(\mathbf{p}_0)
      \mathbf{I})^{i-1}
      \mathbf{Y}^*\frac{\partial \mathbf{A}}{\partial
      p_j}\mathbf{X}\right)-\mathrm{trace}(\mathbf{C}^{i-1})
      \frac{\partial q_1}{\partial p_j}
      -\sum_{k=2}^{i-1}\mathrm{trace}(\mathbf{C}^{i-1}
      \mathbf{E}_{k1})\frac{\partial q_k}{\partial p_j}, \\[15pt]
      i=2,\ldots,d,\quad j=1,\ldots,n,
    \end{array}
  \end{equation}
where the derivatives are evaluated at $\mathbf{p}_0$; $\mathbf{C}
= \mathbf{B}(\mathbf{p}_0)-q_1(\mathbf{p}_0)\mathbf{I}$ is the
companion matrix
    \begin{equation}
    \label{3.7}
    \mathbf{C} = \left(\begin{array}{cccc}
        0 & 1 & & \\
        q_2(\mathbf{p}_0)& 0&\ddots&\\
        \vdots & &\ddots&1\\
        q_d(\mathbf{p}_0)&&&0
    \end{array}\right)
    =\mathbf{J}_0+\sum_{i=2}^d q_i(\mathbf{p}_0)\mathbf{E}_{i1},
    \end{equation}
and $\mathbf{E}_{i1}$ is the matrix having the unit $(i,1)$th
element and zeros in other places.
\end{theorem}

The proof of this theorem is given in the Appendix.

When the matrix $\mathbf{A}$ is arbitrary (not restricted to a
multiparameter matrix family), each entry of the matrix can be
considered is an independent parameter. Hence, the matrix
$\mathbf{A}$ can be used instead of the parameter vector:
$\mathbf{p} \longrightarrow \mathbf{A}$. The derivative of
$\mathbf{A}$ with respect to its $(i,j)$th entry is
$\mathbf{E}_{ij}$. Thus, the formulae of Theorem~\ref{t1} can be
applied.

\begin{corollary}
\label{t3} Let $\mathbf{S}$, $\mathbf{Y}$, and $\mathbf{X}$ be the
matrices satisfying equations (\ref{3.5a}). Then the values of
$q_1(\mathbf{A}_0),\ldots,q_d(\mathbf{A}_0)$ are given by formulae
(\ref{3.6}) and (\ref{3.8}) with $\mathbf{p}_0$ substituted by
$\mathbf{A}_0$. Derivatives of the functions
$q_1(\mathbf{A}),\ldots,q_d(\mathbf{A})$ with respect to
components of the matrix $\mathbf{A}$ taken at $\mathbf{A}_0$ are
    \begin{equation}
    \label{3.24}
    \begin{array}{l}
        \displaystyle
        \frac{\partial q_1}{\partial\mathbf{A}}
        = \left(\mathbf{X}\mathbf{Y}^*/d\right)^T, \\[10pt]
        \displaystyle
        \frac{\partial q_i}{\partial\mathbf{A}}
        = \left(\mathbf{X}(\mathbf{S}-q_1(\mathbf{p}_0)\mathbf{I})^{i-1}
        \mathbf{Y}^*\right)^T-\mathrm{trace}(\mathbf{C}^{i-1})
        \frac{\partial q_1}{\partial\mathbf{A}}
        -\sum_{k=2}^{i-1}\mathrm{trace}
        (\mathbf{C}^{i-1}\mathbf{E}_{k1})
        \frac{\partial q_k}{\partial\mathbf{A}},\\[3pt]
        \qquad\quad i = 2,\ldots,d.
    \end{array}
    \end{equation}
Here $T$ is the transpose operator, and
    \begin{equation}
    \label{3.24b}
    \frac{\partial q_i}{\partial\mathbf{A}}
    = \left(\begin{array}{ccc}
        \displaystyle\frac{\partial q_i}{\partial a_{11}}
        & \cdots &
        \displaystyle\frac{\partial q_i}{\partial a_{1m}} \\[10pt]
        \vdots & \ddots & \vdots \\[3pt]
        \displaystyle\frac{\partial q_i}{\partial a_{m1}}
        & \cdots &
        \displaystyle\frac{\partial q_i}{\partial a_{mm}}
    \end{array}\right)
    \end{equation}
is the $m \times m$ matrix of derivatives of $q_i(\mathbf{A})$
with respect to components of the matrix $\mathbf{A}$ taken at
$\mathbf{A}_0$.
\end{corollary}

At $\mathbf{p}_0 \in \lambda^d$, we can find the multiple
eigenvalue $\lambda$ and the corresponding Jordan chain of
generalized eigenvectors $\mathbf{u}_1,\ldots,\mathbf{u}_d$. This
problem reduces to the transformation of the matrix $\mathbf{S}$
to the prescribed Jordan form (one Jordan block). A possible way
of solving this problem is presented in the following theorem (see
the Appendix for the proof.).

\begin{theorem}
\label{t2} At the point $\mathbf{p}_0 \in \lambda^d$, the multiple
eigenvalue is given by the expression
    \begin{equation}
    \label{3.14}
    \lambda = \mathrm{trace}\,\mathbf{S}/d.
    \end{equation}
The general form of the Jordan chain of generalized eigenvectors
$\mathbf{u}_1,\ldots,\mathbf{u}_d$ is
    \begin{equation}
    \label{3.15}
    \mathbf{u}_1 =
    \mathbf{X}(\mathbf{S}-\lambda\mathbf{I})^{d-1}\mathbf{k},\quad
    \ldots,\quad
    \mathbf{u}_{d-1} =
    \mathbf{X}(\mathbf{S}-\lambda\mathbf{I})\mathbf{k},\ \
    \mathbf{u}_d = \mathbf{X}\mathbf{k},
    \end{equation}
where $\mathbf{k} \in \mathbb{C}^d$ is an arbitrary vector such
that the eigenvector $\mathbf{u}_1$ is nonzero. Choosing a
particular unit-norm eigenvector $\hat{\mathbf{u}}_1$, e.g., by
taking the scaled biggest norm column of the matrix
$\mathbf{X}(\mathbf{S}-\lambda\mathbf{I})^{d-1}$, one can fix the
vector $\mathbf{k}$ by the orthonormality conditions
    \begin{equation}
    \label{3.16}
    \hat{\mathbf{u}}_1^*\mathbf{u}_i
    = \left\{\begin{array}{ll}
        1, & i=1; \\
        0, & i=2,\ldots,d.
    \end{array}\right.
  \end{equation}

The accuracy of the multiple eigenvalue and generalized
eigenvectors determined by formulae (\ref{3.14}) and (\ref{3.15})
has the same order as the accuracy of the point $\mathbf{p}_0$ in
the surface $\lambda^d$.
\end{theorem}

\section{Newton's method}

There are several ways to find the matrices $\mathbf{S}$,
$\mathbf{X}$, and $\mathbf{Y}$. The simplest way is to use the
diagonalization of $\mathbf{A}_0$. Then $\mathbf{S} =
\mathrm{diag}(\lambda_1,\ldots,\lambda_d)$ is the diagonal matrix,
and the columns of $\mathbf{X}$ and $\mathbf{Y}$ are the right and
left eigenvectors corresponding to $\lambda_1,\ldots,\lambda_d$.
This way will be discussed in Section~5.

If the parameter vector $\mathbf{p}_0$ is close to the surface
$\lambda^d$, the diagonalization of the matrix $\mathbf{A}_0$ is
ill-conditioned. Instead of the diagonalization, one can use the
numerically stable Schur decomposition $\widetilde{\mathbf{S}} =
\widetilde{\mathbf{X}}^*\mathbf{A}_0\widetilde{\mathbf{X}}$, where
$\widetilde{\mathbf{S}}$ is an upper-triangular matrix called the
Schur canonical form, and $\widetilde{\mathbf{X}} =
(\widetilde{\mathbf{X}}^*)^{-1}$ is a unitary matrix~\cite{GoVa}.
The diagonal elements
$\widetilde{s}_{11},\ldots,\widetilde{s}_{mm}$ of
$\widetilde{\mathbf{S}}$ are the eigenvalues of $\mathbf{A}_0$. We
can choose the Schur form so that the first $d$ diagonal elements
are the eigenvalues $\lambda_1,\ldots,\lambda_d$. Performing the
block-diagonalization of the Schur form
$\widetilde{\mathbf{S}}$~\cite[\S 7.6]{GoVa}, we obtain the
block-diagonal matrix
    \begin{equation}
    \label{3.4}
    \left(\begin{array}{cc}
        \mathbf{S} & 0 \\ 0 & \mathbf{S}'
    \end{array}\right)
    = [\mathbf{Y}, \mathbf{Y}']^*\mathbf{A}_0
    [\mathbf{X}, \mathbf{X}'],
    \end{equation}
where $\mathbf{S}$ is a $d \times d$ upper-triangular matrix with
the diagonal $(\lambda_1,\ldots,\lambda_d)$;
$[\mathbf{X},\mathbf{X}']$ and
$[\mathbf{Y},\mathbf{Y}']^*=[\mathbf{X},\mathbf{X}']^{-1}$ are
nonsingular $m \times m$ matrices (not necessarily unitary). These
operations with a Schur canonical form are standard and included
in many numerical linear algebra packages, for example,
LAPACK~\cite{LAPACK}. They are numerically stable if the
eigenvalues $\lambda_1,\ldots,\lambda_d$ are separated from the
remaining part of the spectrum. As a result, we obtain the
matrices $\mathbf{S}$, $\mathbf{X}$, and $\mathbf{Y}$ satisfying
equations (\ref{3.5a}).

When the matrices $\mathbf{S}$, $\mathbf{X}$, and $\mathbf{Y}$ are
determined, Theorem~\ref{t1} provides the necessary information
for using Newton's method for determining the stratum $\lambda^d$.
Indeed, having the parameter vector $\mathbf{p}_0 =
(p_1^0,\ldots,p_n^0)$ as the initial guess, we linearize equations
(\ref{2.3}) of the surface $\lambda^d$ as
    \begin{equation}
    \label{3.9}
    q_i(\mathbf{p}_0)+\sum_{j=1}^n \frac{\partial q_i}{\partial p_j}
    (p_j-p_j^0) = 0, \quad
    i = 2,\ldots,d,
    \end{equation}
where the values of $q_i(\mathbf{p}_0)$ and the derivatives
$\partial q_i/\partial p_j$ at $\mathbf{p}_0$ are provided by
Theorem~\ref{t1}. In the generic case, the linear part in
(\ref{3.9}) is given by the maximal rank matrix $[\partial
q_i/\partial p_j]$. System (\ref{3.9}) has the single solution if
the number of parameters $n = d-1$ (the set $\lambda^d$ is an
isolated point). If $n > d-1$, one can take the least squares
solution or any other solution depending on which point of the
surface $\lambda^d$ one would like to find. If $n < d$, the
multiple eigenvalue still can exist in matrices with symmetries
(e.g., Hamiltonian or reversible
matrices~\cite{SeyranianMailybaev2003}); then the least squares
fit solution of (\ref{3.9}) is a good choice.

In Newton's method, the obtained vector of parameters $\mathbf{p}
= (p_1,\ldots,p_n)$ is used in the next iteration. In each
iteration, we should choose $d$ eigenvalues of the matrix
$\mathbf{A}_0$. These are the $d$ eigenvalues nearest to the
approximate multiple eigenvalue
    \begin{equation}
    \label{3.9b}
    \lambda = q_1(\mathbf{p}_0)
    +\sum_{j=1}^n \frac{\partial q_1}{\partial p_j}(p_j-p_j^0)
    \end{equation}
calculated at the previous step. If the iteration procedure
converges, we obtain a point $\mathbf{p} \in \lambda^d$.  Then the
multiple eigenvalue and corresponding generalized eigenvectors are
found by Theorem~\ref{t2}. Note that, at the point $\mathbf{p} \in
\lambda^d$, system (\ref{3.9}) determines the tangent plane to the
surface $\lambda^d$ in parameter space. The pseudo-code of the
described iteration procedure is presented in Table~\ref{tab2}.
Depending on a particular application, the line 3 in this
pseudo-code can be implemented in different ways, e.g., as the
least squares solution or as the solution nearest to the input
parameter vector $\mathbf{p}_0$. The implementation of this method
in MATLAB code is available, see~\cite{MATLABcode}.

\begin{table}
\centering
\begin{flushleft}
    INPUT: matrix family $\mathbf{A}(\mathbf{p})$,
    initial parameter vector $\mathbf{p}_0$, and
    eigenvalues $\lambda_1,\ldots,\lambda_d$
\end{flushleft}
\begin{tabular}{l|l}
  1: & Schur decomposition and
            block-diagonalization (\ref{3.4}) of the matrix
            $\mathbf{A}_0 = \mathbf{A}(\mathbf{p}_0)$;\\[2pt]
  2: & evaluate $q_i(\mathbf{p}_0)$ and
            $\partial q_i/\partial p_j$ by formulae
            (\ref{3.6})--(\ref{3.10});\\[2pt]
  3: & find $\mathbf{p}_{new}$ by
            solving system (\ref{3.9})
            (e.g. the least squares solution);\\[2pt]
  4: & IF $\|\mathbf{p}_{new}-\mathbf{p}_0\|
            > \mathrm{desired\ accuracy}$ \\[2pt]
  5: & \quad evaluate approximate multiple eigenvalue $\lambda_{app}$
            by (\ref{3.9b});\\[2pt]
  6: & \quad choose $d$ eigenvalues
            $\lambda_1^{new},\ldots,\lambda_d^{new}$
            of $\mathbf{A}_{new} = \mathbf{A}(\mathbf{p}_{new})$
            nearest to $\lambda_{app}$;\\[2pt]
  7: & \quad perform a new iteration with
            $\mathbf{p}_0 = \mathbf{p}_{new}$
            and $\lambda_i = \lambda_i^{new}$, $i =
            1,\ldots,d$ (GOTO 1);\\[2pt]
  8: & ELSE (IF $\|\mathbf{p}_{new}-\mathbf{p}_0\|
            \le \mathrm{desired\ accuracy}$) \\[2pt]
  9: & \quad find multiple eigenvalue and generalized eigenvectors
            by formulae (\ref{3.14})--(\ref{3.16});
\end{tabular}
\begin{flushleft}
      OUTPUT: parameter vector $\mathbf{p} \in \lambda^d$,
      multiple eigenvalue $\lambda$ and Jordan chain \\
      \qquad\qquad\quad of generalized eigenvectors
      $\mathbf{u}_1,\ldots,\mathbf{u}_d$
\end{flushleft}
\caption{Pseudo-code of Newton's method for finding multiple
eigenvalues in multiparameter matrix families.}
\label{tab2}
\end{table}

In case of complex matrices dependent on real parameters, the same
formulae can be used. In this case, system (\ref{3.9}) represents
$2(d-1)$ independent equations (each equality determines two
equations for real and imaginary parts). This agrees with the fact
that the codimension of $\lambda^d$ in the space of real
parameters is $2(d-1)$~\cite{Ar2}.

Finally, consider real matrices smoothly dependent on real
parameters. For complex multiple eigenvalues, the system
(\ref{3.9}) contains $2(d-1)$ independent real equations
(codimension of $\lambda^d$ is $2(d-1)$). Remark that imaginary
parts of the eigenvalues $\lambda_1,\ldots,\lambda_d$ should have
the same sign. For real multiple eigenvalues, $q_i(\mathbf{p}_0)$
and $\partial q_i/\partial p_j$ are real (the real Schur
decomposition must be used). Hence, (\ref{3.9}) contains $d-1$
real equations (codimension of $\lambda^d$ is $d-1$). In this
case, the eigenvalues $\lambda_1,\ldots,\lambda_d$ are real or
appear in complex conjugate pairs.

In some applications, like stability
theory~\cite{SeyranianMailybaev2003}, we are interested in
specific multiple eigenvalues, e.g., zero and purely imaginary
eigenvalues. In this case equation (\ref{3.9b}) should be included
in the linear system of Newton's approximation (\ref{3.9}).

For arbitrary matrices $\mathbf{A} = [a_{ij}]$ (without
parameters), similar Newton's iteration procedure is based on
Corollary~\ref{t3}. The linearized equations (\ref{3.9}) are
substituted by
    \begin{equation}
    \label{3.23}
    q_i(\mathbf{p}_0)
    +\sum_{j,k = 1}^m
    \frac{\partial q_i}{\partial a_{jk}}(a_{jk}-a_{jk}^0) = 0,
    \quad i=2,\ldots,d,
    \end{equation}
where $\mathbf{A}_0 = [a_{jk}^0]$ is the matrix obtained at the
previous step or the initial input matrix. The first-order
approximation of the multiple eigenvalue (\ref{3.9b}) takes the
form
    \begin{equation}
    \label{3.25}
    \lambda =     q_1(\mathbf{p}_0)
    +\sum_{j,k = 1}^m
    \frac{\partial q_1}{\partial a_{jk}}(a_{jk}-a_{jk}^0).
    \end{equation}

\section{Examples}

All calculations in the following examples were performed using
MATLAB code~\cite{MATLABcode}. For the sake of brevity, we will
show only first several digits of the computation results.

\subsection{Example 1} 
\begin{figure}
\begin{center}
\includegraphics[angle=0, width=0.5\textwidth]{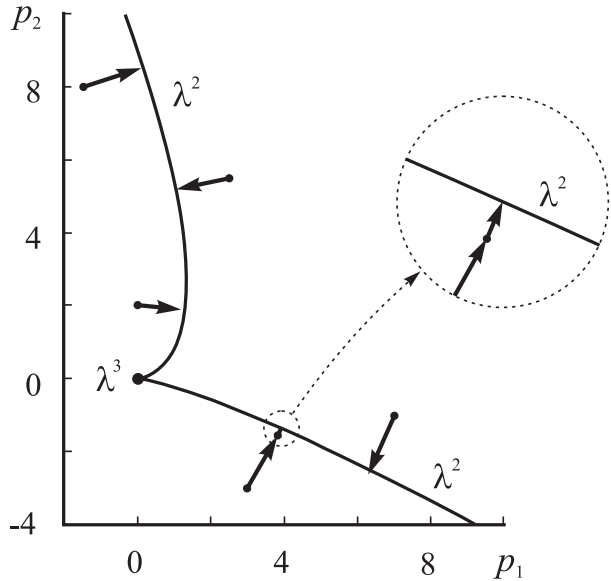}
\end{center}
\caption{One-step approximations of the nearest points with double
eigenvalues.} \label{fig2}
\end{figure}

Let us consider the two-parameter family of
real matrices
  \begin{equation}
  \label{3.30}
  \mathbf{A}(\mathbf{p})=\left(\begin{array}{ccc}
  1 & 3 & 0 \\ p_1 & 1 & p_2 \\ 2 & 3 & 1 \end{array}\right),
  \quad \mathbf{p}=(p_1,p_2).
  \end{equation}
Bifurcation diagram for this matrix family is found analytically
by studying the discriminant of the characteristic polynomial.
There are two smooth curves $\lambda^2$ and a point $\lambda^3$ at
the origin (the cusp singularity), see Figure~\ref{fig2}.

Let us consider the point $\mathbf{p}_0=(-0.03,8.99)$, where the
matrix $\mathbf{A}_0$ has the eigenvalues $\lambda_{1,2}= -1.995
\pm i 0.183$ and $\lambda_3=6.990$. In order to detect a double
real eigenvalue, we choose the pair of complex conjugate
eigenvalues $\lambda_1$, $\lambda_2$. By ordering diagonal blocks
in the real Schur form of $\mathbf{A}_0$ and block-diagonalizing,
we find the matrices $\mathbf{S}$, $\mathbf{X}$, and $\mathbf{Y}$
satisfying (\ref{3.5a}) in the form
  \[
  \mathbf{S}=\left(\begin{array}{cc}
    -1.995 & -5.083 \\
     0.007 & -1.995
  \end{array}\right),\
  \mathbf{X}=\left(\begin{array}{cc}
     0.688 & -0.676 \\
    -0.688 & -0.491 \\
     0.231 &  0.550
   \end{array}\right),\
  \mathbf{Y}=\left(\begin{array}{cc}
     0.729 & -0.574 \\
    -0.604 & -0.286 \\
     0.357 &  0.858
  \end{array}\right).
  \]
Applying the formulae of Theorem~\ref{t1}, we find
  \begin{equation}
  \label{3.32}
  \left(\begin{array}{c}
    q_1(\mathbf{p}_0) \\[2pt]
    q_2(\mathbf{p}_0)
  \end{array}\right)
  = \left(\begin{array}{c}
    -1.995 \\ -0.033
  \end{array}\right),\quad
  \left(\begin{array}{cc}
    \partial q_1/\partial p_1 & \partial q_1/\partial p_2 \\[2pt]
    \partial q_2/\partial p_1 & \partial q_2/\partial p_2
  \end{array}\right)
  = \left(\begin{array}{cc}
  -0.111 & -0.148 \\
   1.001 &  0.333
  \end{array}\right).
  \end{equation}
The linearized system (\ref{3.9}) represents one real scalar
equation. We find the nearest parameter vector $\mathbf{p} \in
\lambda^2$ (the least squares solution) as
  \begin{equation}
  \label{3.33}
  \mathbf{p}
  = \mathbf{p}_0-\frac{q_2(\mathbf{p}_0)}{
  (\partial q_2/\partial p_1)^2+(\partial q_2/\partial p_2)^2}\,
  (\partial q_2/\partial p_1,\, \partial q_2/\partial p_2)
  = (-0.00001, 8.99999).
  \end{equation}
After five iterations of Newton's method, we find the exact
nearest point $\mathbf{p} = (0,\,9) \in \lambda^2$. Then
Theorem~\ref{t2} gives the multiple eigenvalue and the Jordan
chain with the accuracy $10^{-15}$:
  \begin{equation}
  \label{3.35}
  \lambda=-2, \quad
  [\mathbf{u}_1,\mathbf{u}_2] =
  \frac{1}{\sqrt{19}}\left(\begin{array}{cc}
    3 & -1+30/19 \\
    -3 & 2-30/19\\
    1 & -1+10/19
  \end{array}\right).
  \end{equation}

Now let us take different points $\mathbf{p}_0$ in the
neighborhood of the curve $\lambda^2$ and calculate one-step
Newton's approximations of the nearest points $\mathbf{p} \in
\lambda^2$. In this case we choose $\lambda_1,\,\lambda_2$ as a
pair of complex conjugate eigenvalues of
$\mathbf{A}_0=\mathbf{A}(\mathbf{p}_0)$. If all eigenvalues of
$\mathbf{A}_0$ are real, we test all different pairs of
eigenvalues, and take the pair providing the nearest point
$\mathbf{p} \in \lambda^2$. The result is shown in
Figure~\ref{fig2}, where each arrow connects the initial point
$\mathbf{p}_0$ with the one-step Newton's approximation
$\mathbf{p}$. For one point $\mathbf{p}_0$ we performed two
iterations, taking the point $\mathbf{p}$ as a new initial point
$\mathbf{p}_0=\mathbf{p}$. The convergence of this iteration
series is shown in the enlarged part of parameter space (inside
the circle in Figure~\ref{fig2}). The results confirm Newton's
method rate of convergence.

\subsection{Example 2}

Let us consider the real matrix
$\mathbf{A}_0=\mathbf{A}_1+\varepsilon\mathbf{E}$, where
  \begin{equation}
  \label{3.36}
  \mathbf{A}_1=\left(\begin{array}{ccc}
  0 & 1 & 0 \\ 0 & 0 & \delta \\ 0 & 0 & 0
  \end{array}\right), \quad
  \mathbf{E}=\left(\begin{array}{ccc}
  3 & 4 & 2 \\ 8 & 3 & 6 \\ 4 & 9 & 6
  \end{array}\right),
  \end{equation}
and $\varepsilon=\textrm{2.2e-15}$, $\delta=\textrm{1.5e-9}$. This
matrix was used in~\cite{EdMa} for testing the GUPTRI~\cite{KaRu,
KaRu2} algorithm. It turned out that this algorithm detects a
matrix $\mathbf{A} \in \lambda^3$ (with a nonderogatory triple
eigenvalue) at the distance $O(10^{-6})$ from $\mathbf{A}_0$,
while the distance from $\mathbf{A}_0$ to $\lambda^3$ is less than
$\|\varepsilon\mathbf{E}\|_F=\textrm{3.62e-14}$ since
$\mathbf{A}_1 \in \lambda^3$. This is explained by the observation
that the algorithm finds matrix perturbations along a specific set
of directions, and these directions are almost tangent to the
stratum $\lambda^3$ in the case under consideration~\cite{EdMa}.

Our method determines locally the whole stratum $\lambda^3$ in
matrix space and, hence, it should work correctly in this case.
Since the triple eigenvalue is formed by all eigenvalues of
$\mathbf{A}_0$, we can use $\mathbf{S} = \mathbf{A}_0$ and
$\mathbf{X} = \mathbf{Y} = \mathbf{I}$ in the formulae of
Corollary~\ref{t3}. As a result, we find the least squares
solution of system (\ref{3.23}) in the form
$\mathbf{A}=\mathbf{A}_0+\Delta\mathbf{A}$, where
  \begin{equation}
  \label{3.37}
  \Delta\mathbf{A}=\textrm{1.0e-14}*\left(\begin{array}{ccc}
    0 & 0 & 0 \\
    -1.760 & 0 &  0 \\
    -0.880 & 0 &  0
  \end{array}\right).
  \end{equation}
Approximations of the multiple eigenvalue and corresponding
generalized eigenvectors evaluated by Theorem~\ref{t2} for the
matrix $\mathbf{A}$ are
  \begin{equation}
  \label{3.38}
    \lambda=\textrm{8.800e-15}, \quad
    [\mathbf{u}_1,\,\mathbf{u}_2,\,\mathbf{u}_3]
    = \left(\begin{array}{ccc}
    1.000 & -0.000 & -0.000 \\
    0.000 & 1.000 & -0.000 \\
    0.000 & 0.000 & \textrm{6.667e+8}
  \end{array}\right).
  \end{equation}

We detected the matrix $\mathbf{A}$ at the distance
$\|\Delta\mathbf{A}\|_F=\textrm{1.97e-14}$, which is smaller than
the initial perturbation
$\|\varepsilon\mathbf{E}\|_F=\textrm{3.62e-14}$
($\|\Delta\mathbf{A}\|_F$ denotes the Frobenius matrix norm). The
matrix $\mathbf{U} = [\mathbf{u}_1,\mathbf{u}_2,\mathbf{u}_3]$
satisfies the Jordan chain equation (\ref{3.2}) with the very high
accuracy $\|\mathbf{A}\mathbf{U}-\mathbf{U}\mathbf{J}_\lambda\|_F
/\|\mathbf{U}\|_F=\textrm{9.6e-23}$.

The normal complementary subspace $N$ of the tangent space to
$\lambda^3$ at $\mathbf{A}_1$ has the form~\cite{Ar2}
  \begin{equation}
  \label{3.39}
    N=\bigg\{\left(\begin{array}{ccc}
    0 & 0 & 0 \\
    x & 0 & 0 \\
    y & \delta x & 0
  \end{array}\right) \mid \ x,y\in\mathbb{R}\bigg\}.
  \end{equation}
It is easy to see that the matrix $\Delta\mathbf{A}$ in
(\ref{3.37}) is equal to the projection of
$\,-\varepsilon\mathbf{E}$ to the normal subspace $N$. This
confirms that the obtained matrix $\mathbf{A} \in \lambda^3$ is
the nearest to $\mathbf{A}_0$.

\subsection{Example 3}

Let us consider the $12 \times 12$ Frank matrix $\mathbf{A}_0 =
[a_{ij}^0]$ with the elements
  \begin{equation}
  \label{3.45}
  a_{ij}^0 = \left\{\begin{array}{ll}
    n+1-\max(i,j), & j\ge i-1, \\
    0, & j<i-1.
  \end{array}\right.
  \end{equation}
The Frank matrix has six small positive eigenvalues which are
ill-conditioned and form nonderogatory multiple eigenvalues of
multiplicities $d=2,\ldots,6$ for small perturbations of the
matrix. The results obtained by Newton's method with the use of
Corollary~\ref{t3} are presented in Table~\ref{tab1}. An
eigenvalue of multiplicity $d$ of the nearest matrix $\mathbf{A}
\in \lambda^d$ is formed by $d$ smallest eigenvalues of
$\mathbf{A}_0$. The second column of Table~\ref{tab1} gives the
distance $\mathrm{dist}(\mathbf{A}_0,\lambda^d) =
\|\mathbf{A}-\mathbf{A}_0\|_F$, where the matrix $\mathbf{A}$ is
computed after one step of Newton's procedure. The third column
provides exact distances computed by Newton's method, which
requires 4--5 iterations to find the distance with the accuracy
$O(10^{-15})$. At each iteration, we find the solution
$\mathbf{A}$ of system (\ref{3.23}), which is the nearest to the
matrix (\ref{3.45}). The multiple eigenvalues and corresponding
generalized eigenvectors are found at the last iteration by
Theorem~\ref{t2}. The accuracy estimated as
$\|\mathbf{A}\mathbf{U} -\mathbf{U}\mathbf{J}_{\lambda}\|_F
/\|\mathbf{U}\|_F$ varies between $10^{-10}$ and $10^{-13}$. The
matrices of generalized eigenvectors $\mathbf{U}$ have small
condition numbers, which are given in the fourth column of
Table~\ref{tab1}. For comparison, the fifth and sixth columns give
upper bounds for the distance to the nearest matrix
$\mathbf{A}\in\lambda^d$ found in~\cite{Fa, KaRu}.

\begin{table}
\begin{center}
\begin{small}
\begin{tabular}{|c|c|c|c|c|c|}
\hline
  $d$ & $\begin{array}{c} \textrm{dist}(\mathbf{A}_0,\lambda^d) \\
        \textrm{1-step approximation}\end{array}$ &
  $\begin{array}{c} \textrm{dist}(\mathbf{A}_0,\lambda^d) \\
        \textrm{exact} \end{array} $ &
    {\rm cond}\,$\mathbf{U}$ &
    $\|\mathbf{A}-\mathbf{A}_0\|_F$ \cite{Fa} & $\|\mathbf{A}-\mathbf{A}_0\|_F$ \cite{KaRu} \\ \hline
  2 & \textrm{1.619e-10} & \textrm{1.850e-10} & \textrm{1.125} & \textrm{3.682e-10} & \textrm{} \\ \hline
  3 & \textrm{1.956e-8} & \textrm{2.267e-8} & \textrm{1.746} & \textrm{3.833e-8} & \textrm{} \\ \hline
  4 & \textrm{1.647e-6} & \textrm{1.861e-6} & \textrm{4.353} & \textrm{3.900e-6} & \textrm{} \\ \hline
  5 & \textrm{9.299e-5} & \textrm{1.020e-4} & \textrm{14.14} & \textrm{4.280e-4} & \textrm{6e-3} \\ \hline
  6 & \textrm{3.150e-3} & \textrm{3.400e-3} & \textrm{56.02} & \textrm{7.338e-2} & \textrm{} \\
\hline
\end{tabular}
\end{small}
\end{center}
\caption{Distances to the multiple eigenvalue strata $\lambda^d$
for the Frank matrix.} \label{tab1}
\end{table}

We emphasize that this is the first numerical method that is able
to find exact distance to a nonderogatory stratum $\lambda^d$.
Methods available in the literature cannot solve this problem
neither in matrix space nor for multiparameter matrix families.

\section{Convergence and accuracy}

In the proposed approach, the standard Schur decomposition and
block-diagonalization (\ref{3.4}) of a matrix are required at each
iteration step. Additionally, first derivatives of the matrix with
respect to parameters are needed at each step. Numerical accuracy
of the block-diagonalization depends on the separation ${\rm
sep}(\mathbf{S},\mathbf{S}')$ of the diagonal blocks in the Schur
canonical form (calculated prior the
block-diagonalization)~\cite{GoVa}. Instability occurs for very
small values of ${\rm sep}(\mathbf{S},\mathbf{S}')$, which
indicates that the spectra of $\mathbf{S}$ and $\mathbf{S}'$
overlap under a very small perturbation of $\mathbf{A}_0$. Thus,
numerical instability signals that the chosen set of $d$
eigenvalues should be changed such that the matrix $\mathbf{S}$
includes all "interacting" eigenvalues.

The functions $q_i(\mathbf{p})$ are strongly nonlinear near the
boundary of the surface $\lambda^d$. The boundary corresponds to
higher codimension strata associated with eigenvalues of higher
multiplicity (or eigenvalues of the same multiplicity but with
several Jordan blocks). For example, the stratum $\lambda^2$ in
Figure~\ref{fig1} is bounded by the singularities $\lambda^3$ and
$\lambda^4$. As a result, the convergence of Newton's method may
be poor near the boundary of $\lambda^d$. This instability signals
that we should look for eigenvalues with a more degenerate Jordan
structure (e.g. higher multiplicity $d$). Analysis of the surface
$\lambda^d$ very close to the boundary is still possible, but the
higher precision arithmetics may be necessary.

\begin{figure}
\begin{center}
\includegraphics[angle=0, width=0.5\textwidth]{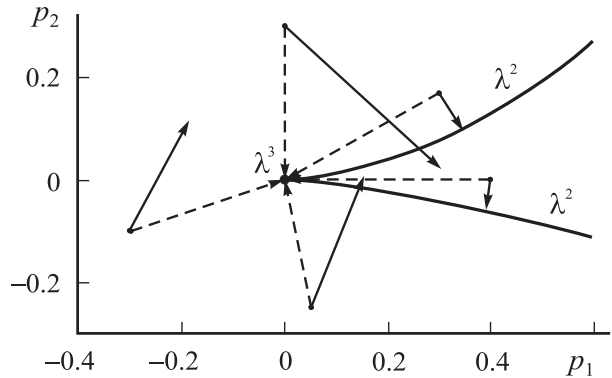}
\end{center}
\caption{One-step approximations of the nearest points of the
strata $\lambda^2$ (solid arrows) and $\lambda^3$ (dashed arrows)
near the cusp singularity.} \label{fig3}
\end{figure}

Figure~\ref{fig3} shows first iterations of Newton's procedure for
different initial points in parameter space for matrix family
(\ref{3.30}) from Example~1. Solid arrows locate double
eigenvalues (the stratum $\lambda^2$) and the dashed arrows
correspond to triple eigenvalues (the stratum $\lambda^3$). One
can see that the stratum $\lambda^2$ is well approximated when
$\mathbf{p}_0$ is relatively far from the singularity (from the
more degenerate stratum $\lambda^3$). For the left-most point
$\mathbf{p}_0$ in Figure~\ref{fig3}, the nearest point
$\mathbf{p}\in\lambda^2$ simply does not exist (infimum of the
distance $\|\mathbf{p}-\mathbf{p}_0\|$ for
$\mathbf{p}\in\lambda^2$ corresponds to the origin
$\mathbf{p}=0\in\lambda^3$). Note that, having the information on
the stratum $\lambda^3$, it is possible to determine locally the
bifurcation diagram (describe the geometry of the cusp singularity
in parameter space)~\cite{Ma, SeyranianMailybaev2003}.

For the backward error analysis of numerical eigenvalue problems
based on the study of the pseudo-spectrum we refer
to~\cite{Traviesas2000}. We note that the classical numerical
eigenvalue problem is ill-conditioned in the presence of multiple
eigenvalues. The reason for that is the nonsmoothness of
eigenvalues at multiple points giving rise to singular
perturbation terms of order $\varepsilon^{1/d}$, where $d$ is the
size of Jordan block~\cite{SeyranianMailybaev2003}. On the
contrary, in our problem we deal with the regular smooth objects:
the strata $\lambda^d$ and the versal deformation
$\mathbf{B}(\mathbf{p})$.

\section{Approximations based on diagonal decomposition}

In this section we consider approximations derived by using the
diagonal decomposition of $\mathbf{A}_0$. The diagonal
decomposition is known to be ill-conditioned for nearly defective
matrices. However, this way is easy to implement, while the very
high accuracy may be not necessary. According to bifurcation
theory for eigenvalues~\cite{SeyranianMailybaev2003}, the accuracy
of the results based on the diagonal decomposition will be of
order $\varepsilon^{1/d}$, where $\varepsilon$ is the arithmetics
precision. Another reason is theoretical. Bifurcation theory
describes the collapse of a Jordan block into simple
eigenvalues~\cite{MoBuOv, SeyranianMailybaev2003, ViLy}. Our
approximations based on the diagonal decomposition solve the
inverse problem: using simple (perturbed) eigenvalues and
corresponding eigenvectors, we approximate the stratum $\lambda^d$
at which these eigenvalues coalesce.

Let us assume that the matrix $\mathbf{A}_0$ is diagonalizable
(its eigenvalues $\lambda_1,\ldots,\lambda_m$ are distinct). The
right and left eigenvectors of $\mathbf{A}_0$ are determined by
the equations
  \begin{equation}
  \label{6.2}
    \mathbf{A}_0\mathbf{x}_i=\lambda_i\mathbf{x}_i,\quad
    \mathbf{y}_i^*\mathbf{A}_0=\lambda_i\mathbf{y}_i^*,\quad
    \mathbf{y}_i^*\mathbf{x}_i=1
  \end{equation}
with the last equality being the normalization condition.

In Theorem~\ref{t1} we take $\mathbf{S}={\rm
diag}(\lambda_1,\ldots,\lambda_d)=\mathbf{Y}^*\mathbf{A}_0\mathbf{X}$,
$\mathbf{X}=[\mathbf{x}_1,\ldots,\mathbf{x}_d]$, and
$\mathbf{Y}=[\mathbf{y}_1,\ldots,\mathbf{y}_d]$, where
$\lambda_1,\ldots,\lambda_d$ are the eigenvalues coalescing at
$\mathbf{p}\in\lambda^d$. In this case expressions (\ref{3.10})
take the form
    \begin{equation}
    \label{6.3}
    \begin{array}{l}
        \displaystyle
        \frac{\partial q_1}{\partial p_j} = \frac{1}{d}\sum_{i=1}^d
        \mathbf{y}_i^*\frac{\partial \mathbf{A}}{\partial
        p_j}\mathbf{x}_i, \\[15pt]
        \displaystyle
        \frac{\partial q_i}{\partial p_j}
        = \sum_{i=1}^d (\lambda_i-q_1(\mathbf{p}_0))^{i-1}
        \mathbf{y}_i^*\frac{\partial \mathbf{A}}{\partial
        p_j}\mathbf{x}_i-\mathrm{trace}(\mathbf{C}^{i-1})
        \frac{\partial q_1}{\partial p_j}
        -\sum_{k=2}^{i-1}\mathrm{trace}(\mathbf{C}^{i-1}
        \mathbf{E}_{k1})\frac{\partial q_k}{\partial p_j}, \\[15pt]
        \qquad i=2,\ldots,d;\ j=1,\ldots,n.
    \end{array}
    \end{equation}
The interesting feature of these expressions is that they depend
only on the simple eigenvalues $\lambda_1,\ldots,\lambda_d$ and
their derivatives with respect to parameters at
$\mathbf{p}_0$~\cite{SeyranianMailybaev2003}:
    \begin{equation}
    \label{6.4}
    \frac{\partial \lambda_i}{\partial p_j}
    =\mathbf{y}_i^*\frac{\partial \mathbf{A}}{\partial p_j}\mathbf{x}_i, \quad
    i=1,\ldots,d.
    \end{equation}

For example, for $d = 2$ we obtain the first-order approximation
of the surface $\lambda^2$ in the form of one linear equation
    \begin{equation}
    \label{6.5}
    \delta^2+\delta\sum_{j=1}^n
    \left(\mathbf{y}_2^*\frac{\partial \mathbf{A}}{\partial p_j}\mathbf{x}_2
    -\mathbf{y}_1^*\frac{\partial \mathbf{A}}{\partial
    p_j}\mathbf{x}_1\right)(p_j-p_j^0)=0, \quad
    \delta=\frac{\lambda_2-\lambda_1}{2}.
    \end{equation}
Let us introduce the gradient vectors $\nabla \lambda_i=(\partial
\lambda_i/\partial p_1, \ldots, \partial \lambda_i/\partial p_n)$,
$i=1,2$, with the derivatives $\partial \lambda_i/\partial p_j$
given by expression (\ref{6.4}). Then the solution of (\ref{6.5})
approximating the vector $\mathbf{p}\in\lambda^2$ nearest to
$\mathbf{p}_0$ is found as
    \begin{equation}
    \label{6.6}
    \mathbf{p}=\mathbf{p}_0-\frac{\overline{\nabla \lambda_2}
    -\overline{\nabla \lambda_1}}{\|\nabla \lambda_2-\nabla
    \lambda_1\|^2}\,\delta.
    \end{equation}

It is instructive to compare this result with the first-order
approximation of the nearest $\mathbf{p}\in\lambda^2$, if we
consider $\lambda_1$ and $\lambda_2$ as smooth functions
    \begin{equation}
    \label{6.7}
    \lambda_i(\mathbf{p})=\lambda_i+\sum_{j = 1}^n
    \frac{\partial\lambda_i}{\partial p_j}\,
    (p_j-p_j^0)+O(\|\mathbf{p}-\mathbf{p}_0\|^2), \quad i=1,2.
    \end{equation}
Using (\ref{6.7}) in the equation
$\lambda_1(\mathbf{p})=\lambda_2(\mathbf{p})$ and neglecting
higher order terms, we find
    \begin{equation}
    \label{6.8}
    \sum_{j = 1}^n \left(
    \frac{\partial\lambda_2}{\partial p_j}
    -\frac{\partial\lambda_1}{\partial p_j}\right)
    (p_j-p_j^0) = \lambda_1-\lambda_2,
    \end{equation}
which yields the nearest $\mathbf{p}$ as
    \begin{equation}
    \label{6.9}
    \mathbf{p}=\mathbf{p}_0-\frac{\overline{\nabla \lambda_2}
    -\overline{\nabla \lambda_1}}{\|\nabla \lambda_2-\nabla
    \lambda_1\|^2}\,2\delta.
    \end{equation}

Comparing (\ref{6.9}) with (\ref{6.6}), we see that considering
simple eigenvalues as smooth functions, we find the correct
direction to the nearest point $\mathbf{p}\in\lambda^2$, but make
a mistake in the distance to the stratum $\lambda^2$
overestimating it exactly twice. This is the consequence of the
bifurcation taking place at $\mathbf{p}\in\lambda^2$ and resulting
in $O(\|\mathbf{p}-\mathbf{p}_0\|^{1/2})$ perturbation of
eigenvalues and eigenvectors~\cite{MoBuOv, SeyranianMailybaev2003,
ViLy}.

\section{Conclusion}

In the paper, we developed Newton's method for finding multiple
eigenvalues with one Jordan block in multiparameter matrix
families. The method provides the nearest parameter vector with a
matrix possessing an eigenvalue of given multiplicity. It
also gives the generalized eigenvectors and describes the
local structure (tangent plane) of the stratum $\lambda^d$. The
motivation of the problem comes from applications, where matrices
describe behavior of a system depending on several parameters.

The whole matrix space has been considered as a particular case,
when all entries of a matrix are independent parameters. Then the
method provides an algorithm for solving the Wilkinson problem of
finding the distance to the nearest degenerate matrix.

Only multiple eigenvalues with one Jordan block have been studied.
Note that the versal deformation is not universal for multiple
eigenvalues with several Jordan blocks (the functions
$q_1(\mathbf{p}),\ldots,q_d(\mathbf{p})$ are not uniquely
determined by the matrix family)~\cite{Ar1, Ar2}. This requires
modification of the method. Analysis of this case is the topic for
further investigation.

\section{Appendix}

\subsection{Proof of Theorem \ref{t1}}

Taking equation (\ref{2.4}) at $\mathbf{p}_0$, we obtain
    \begin{equation}
    \label{4.10}
    \mathbf{A}_0\mathbf{U}_0
    = \mathbf{U}_0\mathbf{B}_0,
    \end{equation}
where $\mathbf{U}_0 = \mathbf{U}(\mathbf{p}_0)$ and $\mathbf{B}_0
= \mathbf{B}(\mathbf{p}_0)$. Comparing (\ref{4.10}) with
(\ref{3.5a}), we find that the matrix $\mathbf{B}_0$ is equivalent
up to a change of basis to the matrix $\mathbf{S}$. Then the
equality (\ref{3.6}) is obtained by equating the traces of the
matrices $\mathbf{B}_0$ and $\mathbf{S}$, where $\mathbf{B}_0$ has
the form (\ref{2.4}). Similarly, the equality (\ref{3.8}) is
obtained by equating the characteristic equations of the matrices
$\mathbf{B}_0-q_1(\mathbf{p}_0)\mathbf{I}$ and
$\mathbf{S}-q_1(\mathbf{p}_0)\mathbf{I}$.

The columns of the matrices $\mathbf{X}$ and $\mathbf{U}_0$ span
the same invariant subspace of $\mathbf{A}_0$. Hence, the matrices
$\mathbf{X}$ and $\mathbf{U}_0$ are related by the expression
  \begin{equation}
  \label{4.12}
    \mathbf{U}_0=\mathbf{X}\mathbf{F},
  \end{equation}
for some nonsingular $d\times d$ matrix $\mathbf{F}$. Using
(\ref{3.5a}) and (\ref{4.12}) in (\ref{4.10}), we find the
relation
  \begin{equation}
  \label{4.13}
    \mathbf{S}\mathbf{F} = \mathbf{F}\mathbf{B}_0.
  \end{equation}

Taking derivative of equation (\ref{2.4}) with respect to
parameter $p_j$ at $\mathbf{p}_0$, we obtain
    \begin{equation}
    \label{4.11}
    \mathbf{A}_0\frac{\partial \mathbf{U}}{\partial p_j}
    -\frac{\partial \mathbf{U}}{\partial p_j}\mathbf{B}_0
    = \mathbf{U}_0 \frac{\partial \mathbf{B}}{\partial p_j}
    -\frac{\partial \mathbf{A}}{\partial p_j}\mathbf{U}_0.
    \end{equation}
Let us multiply both sides of (\ref{4.11}) by the matrix
$\mathbf{F}^{-1}(\mathbf{S}-q_1(\mathbf{p}_0)
    \mathbf{I})^{i-1}\mathbf{Y}^*$ and take the trace. Using expressions
(\ref{3.5a}), (\ref{4.13}), and the property
$\mathrm{trace}(\mathbf{A}\mathbf{B}) =
\mathrm{trace}(\mathbf{B}\mathbf{A})$, it is straightforward to
check that the left-hand side vanishes and we obtain the equation
    \begin{equation}
    \label{4.18}
    0
    =\mathrm{trace}\left(\Big(\mathbf{U}_0
    \frac{\partial \mathbf{B}}{\partial p_j}
    -\frac{\partial \mathbf{A}}{\partial p_j}\mathbf{U}_0\Big)
    \mathbf{F}^{-1}(\mathbf{S}-q_1(\mathbf{p}_0)
    \mathbf{I})^{i-1}\mathbf{Y}^*\right).
    \end{equation}
Substituting (\ref{4.12}) into (\ref{4.18}) and using equalities
(\ref{3.5a}), (\ref{4.13}), and $\mathbf{B}_0 =
q_1(\mathbf{p}_0)\mathbf{I}+\mathbf{C}$, we find
  \begin{equation}
  \label{4.19}
    \mathrm{trace}\left(\mathbf{C}^{i-1}
    \frac{\partial \mathbf{B}}{\partial p_j}\right)
    -\mathrm{trace}\left((\mathbf{S}-q_1(\mathbf{p}_0)
    \mathbf{I})^{i-1}\mathbf{Y}^*\frac{\partial
    \mathbf{A}}{\partial p_j}\mathbf{X}\right)
    =0.
  \end{equation}
Using (\ref{2.4}), (\ref{3.7}) and taking into account that
$\mathrm{trace}(\mathbf{C}^{i-1}\mathbf{E}_{k1})=0$ for $k>i$, we
obtain
  \begin{equation}
  \label{4.20}
    \mathrm{trace}(\mathbf{C}^{i-1})
    \frac{\partial q_1}{\partial p_j}
    +\sum_{k=2}^i \mathrm{trace}(\mathbf{C}^{i-1}
    \mathbf{E}_{k1})\frac{\partial q_k}{\partial p_j}
    =\mathrm{trace}\left((\mathbf{S}-q_1(\mathbf{p}_0)
    \mathbf{I})^{i-1}\mathbf{Y}^*\frac{\partial
    \mathbf{A}}{\partial p_j}\mathbf{X}\right).
  \end{equation}
Taking equation (\ref{4.20}) for $i=1,\ldots,d$ and using the
equality $\mathrm{trace}(\mathbf{C}^{i-1}\mathbf{E}_{i1})=1$, we
get the recurrent procedure (\ref{3.10}) for calculation of
derivatives of $q_1(\mathbf{p}),\ldots, q_d(\mathbf{p})$ at
$\mathbf{p}_0$.

\subsection{Proof of Theorem \ref{t2}}

At $\mathbf{p}_0 \in \lambda^d$, we have $q_2(\mathbf{p}_0) =
\cdots = q_d(\mathbf{p}_0) = 0$. From (\ref{2.4}) it follows that
$\mathbf{B}_0 = \mathbf{J}_\lambda$ is the Jordan block with the
eigenvalue $\lambda = q_1(\mathbf{p}_0) =
\mathrm{trace}\,\mathbf{S}/d$. By using (\ref{3.5a}), one can
check that the vectors (\ref{3.15}) satisfy the Jordan chain
equations (\ref{3.1}) for any vector $\mathbf{k}$. Equations
(\ref{3.16}) provide the way of choosing a particular value of the
vector $\mathbf{k}$.

Since $\mathbf{B}_0 = \mathbf{J}_\lambda$, the versal deformation
equation (\ref{2.4}) becomes the Jordan chain equation (\ref{3.2})
at $\mathbf{p}_0 \in \lambda^d$. Hence, the columns of the matrix
$\mathbf{U}_0$ are the generalized eigenvectors. Since the
function $q_1(\mathbf{p})$ and the matrix $\mathbf{U}(\mathbf{p})$
smoothly depend on parameters, the accuracy of the multiple
eigenvalue and generalized eigenvectors has the same order as the
accuracy of $\mathbf{p}_0 \in \lambda^d$.

\section*{Acknowledgments}
The author thanks A. P. Seyranian and Yu. M. Nechepurenko for
fruitful discussions. This work was supported by the research
grants RFBR 03-01-00161, CRDF-BRHE Y1-M-06-03, and the President
of RF grant MK-3317.2004.1.

\end{document}